X-ray thermal emission from the jet of M87 with *Chandra*

S.Osone
Funabashi, Chiba, Japan
osonesatoko@gmail.com

Abstract
With new calibration data, thermal emission from the jet of radio galaxy M87 is studied with about 700 ks archival data with *Chandra*. For nucleus, HST-1, knot D, X-ray energy spectra is well fitted with a power law. However, For knot A, a power law model is rejected with a high significance and an X-ray energy spectra is well fitted with a combination model of a power law and an apec model of 0.2 keV and a metal abundance 0.00. Thermal emission from knot A is confirmed.

1.Introduction

M87 is radio galaxy and is center of Virgo cluster. The distance is 16 Mpc (z=0.004) (Tonry 1991). M87 is famous with Blackhole shadow by Event Horizon Telescope(EHT) observation and a mass of Blackhole is estimated as $(6.5\pm0.7)\ 10^9 M_\odot$(EHT2019).

M87 is observed with multiwave lengths from radio to TeV gamma ray. M87 has an inclined jet of 20 arc second length. An image resolution is micro arc second in radio, 0.7 arc second in optical and 0.5 arc second in Xray with *Chandra*, 5.2 arc minutes in GeV gamma ray with *Fermi* and 6 arc minutes in TeV gamma ray. Therefore, in high energy gamma ray, the jet cannot be resolved. Therefore, the origin of gamma ray is studied with both an estimated size from flux time variability and a flux correlation of a simultaneous observation with multiwave lengths. A time scale of flux variability is 2d in TeV gamma ray (Aharonian et al. 2006). The size of emitting area is $ct_{val} = 3.1 \times 10^{16}$ (δ/6) cm. Here, c is a speed of light and δ is a doppler factor. Therefore, the origin of TeV gamma ray is considered as nucleus. A flux correlation between nucleus in radio band and TeV gamma ray is reported (Acciari et al. 2009). A flux correlation between nucleus in X-ray band and TeV gamma ray in 2008 and 2010, that between HST-1 in X-ray and TeV gamma ray in 2005 are reported (Abramowaski et al. 2012). Therefore, HST-1 is also considered as the origin of TeV gamma ray.

Synchrotron self compton (SSC) model of accelerated electrons is often used for a multiwave lengths spectra. There is also hybrid model of accelerated electrons and accelerated protons in the jet (Acciari et al. 2020, Alfaro et al. 2022). In a hybrid model, SSC model is used from radio to GeV gamma ray and synchrotron emission of accelerated protons is used in TeV gamma ray. There is flux time variability from radio to TeV gamma ray. However, flux time variability in GeV gamma ray has not been observed. A chance probability of non flux time variability is 0.018 below 10 GeV and 0.23 above 10 GeV with *Fermi* data set from 2008 to 2016 (Benkhali et al. 2019). GeV gamma ray may be a different origin. Osone(2017) calculated non thermal bremsstrahlung flux of accelerated electrons from HST-1 with an X-ray spectral analysis and calculated flux is comparable with *Fermi* flux and an energy loss time scale of non thermal bremsstrahlung from HST-1 is $8.4 \times 10^7 (6/\delta)$ yr. This timescale matches with non flux time variability.

For a spectral analysis, a large exposure time is needed. Because poor statistics leads that any model is acceptable and a subtraction of background is insufficient. Therefore, Osone(2017) and Sun et al(2018) analyzed an X-ray energy spectra of the M87 jet with *Chandra* by using a plenty of an archival data. Osone(2017) reported there is a soft excess against a power law for knot A. Sun et al (2018) reported no soft excess for knot A. For almost same data set, there is a different result. There is a difference in a background region. Sun et al.(2018) use a common region along the jet as a background for all knots. As a background for a jet analysis, hot gas of Virgo cluster and cosmic X ray background and non X ray background of a detector are considered. The X ray emission of hot gas of cluster depend on a distance from the nucleus (Bohringer et al. 2001). Osone(2017) use two neighbour regions as a background for each knot.

Osone(2017) use calibration data of caldb 4.6.8 for data set from 2000 to 2014 and caldb 4.7.6 for data set from 2015 to 2018. Sun et al (2018) use calibration data of caldb 4.7.2 and analyze data set from 2000 to 2016/6. An incorrectness of calibration data is reported by Plucinsky et al.(2018). The caldb 4.7.8 is reported that there is well calibrated from 2000 to 2016 and there is overestimating absorption from 2016 (Plucinksky et al. 2018). Osone(2017) and Sun et al.(2018) use an older version of calibration data than caldb 4.7.8. In this paper, for same data set from 2000 to 2018, by using new calibration data of caldb 4.10.2 which obtained at 2022/12, an energy spectra of the jet of M87 is studied. Thermal emission is confirmed for the knot A with a high significance.

2. observation

*Chandra* has the highest angular resolution of 0.5 arc second among X-ray satellite and resolve the jet, nucleus, HST-1, knot D and knot A. The detector used is CCD. The image is taken from 2-dimension arrays of CCD and an X-ray energy spectra is taken from a deposit energy in CCD. The frame time of 0.4 sec is used. The frame time of 3.2 sec is saturated for these knots. The archival data from 2000 to 2018 is used, which is same data with Osone(2017). I call data from 2000 to 2014 as a former data set, data from

2015 to 2018 as a latter data set as Osone(2017).

3. data analysis

3.1 image

CIAO software of 4.15 and ds9 v8.4 are used. The image is smoothed with a gaussian of 1 sigma=0.5 arc second and displayed with a log scale. The region file which show a position of a source and its radius and that which show a position of background and its radius are made with ds9. The radius of circle as source region and a background region is 0.5 arc second for nucleus, 0.6 arc second for HST-1, 0.75 arcsecond for knot D and 1 arc second for knot A. The background region is a neighbour to source region, north and south of same radius and same distance from nucleus for each part of the jet. A position of a source region for obsID 18232 is shown in table1. The exposure time of obsID 18232 is 18 ks. The image of obsID18232 is shown in figure1.

A position of each knot is changed from data to data. A position of an extracted region is decided by image for each data set. A radius of an extracted region is same by a data set. HST-1 and nucleus sometimes cannot be resolved. These data set are not used.

3.2 pile up treatment

Nucleus and HST-1 are sometimes piled up heavily in a former data set. Heavy pile up distorts X-ray energy spectra. In addition to a calibration data, there are two revised things against Osone(2017) about a pile up. One is a removal data set with a pileup line on image. The other is a change of a tool of a pileup check.

The pile up line is checked with an image. An image is made with a zoom of 4 and a log scale. The pile up line is originated from nucleus or HST-1. The pile up line is sometimes overlapped with a source region or a background region for knot D and knot A. Therefore, data set of not only HST-1 and nucleus but also knot D and knot A are not used.

The degree of a pile up is checked by a pileup_map command for data set which pass a pile up line check. This command makes a pile up information from an event data and the count/frame is calculated for a given region file by a dmstat command. The PIMMS tool used in Osone(2017) is a proposal tool and not an accurate tool of a pile up check.

By a pile up selection, there is small exposure time for nucleus and HST-1 for a former data set. Therefore, two kinds of data set, less than a 5% pileup contamination and less than a 10% pile up contamination are prepared for nucleus and HST-1. For a latter data set, all data is less than a 5% pileup contamination. The used observation log is shown in table 2, for nucleus, HST-1, knot D and knot A, respectively.

3.2 energy spectra

An energy spectra of source region and background region are made with a specextract command from fits data and a region file, respectively. The effective area (arf) is made with no weight of a count rate and a correction by PSF which are suitable conditions for a point source analysis. Arf and an energy response(rmf) are made for only source region. Arf and rmf are multiplied as rsp for each data set and rsp for summed data set is weighted with an exposure time. An area of background region is two times as large as that of source region. When an energy spectra is made, a keyword BACKSCAL in a spectra file for background is set two time as large as that for source. Data points in energy spectra of a source are binned so that minimun counts per a bin are above 15.

The summed energy spectra which subtract background is described with no line feature and a contamination of hot gas of cluster is excluded successfully.

3.2 model fitting

CCD has a sensitivity from 0.2 keV to 10 keV. There is a quantum efficiency degradation by a contamination of an optical filter. An energy range above 0.3 keV, especially above 0.5 keV is well calibrated. Therefore, a lower limit of an energy in energy spectra is set as 0.5 keV.

XSPEC tool is used for an energy spectra analysis. At first, a power law model as synchrotron emission of accelerated electrons is used. Here, an absorption in soft X-ray is caused by a photo electric effect of a neutral material in a line of a sight from us to M87. The column density by a 21 cm radio observation is $1.6 \times 10^{-20} cm^{-2}$ (Kalberia et al 2005). An absorption is applied to a power law. At first, a column density is set free. When a column density is lower than the 21 cm value, a column density is fixed to the 21 cm value and energy spectra is fitted with a power law. When a power law model is not reasonable, an apec model as thermal bremsstrahlung with a metal is added to a power law and energy spectra is fitted. Here, an absorption is also applied to an apec model.

4.result

4.1 former data set

Fitting results are shown in table 3 and 4. For nucleus and HST-1 of two kinds of pile up contamination, when an energy spectra is well fitted with a power law model. For knot D, when an energy spectra is fitted with a power law, the column density is lower than the 21 cm value. Therefore, a column density is fixed to the 21 cm value and energy spectra is well fitted with a power law. For knot A, when an energy spectra is fitted with a power law, a column density is lower than the 21cm value. Therefore, a column density is fixed to the 21 cm value, and energy spectra is fitted with power law. The chance probability of power law is $1.14 \times 10^{-11}$. There is soft excess. Therefore, an apec model is added to a power law and a reasonable result is obtained. The temperature of 0.2 keV and metal abundance of 0.00 are obtained. In an apec model, a relative abundance is fixed to a value of Grevesse & Anders (1989). A relative abundance is cannot be free because of large uncertainty.

4.2 latter data set

Fitting results are shown in table 5 and 6. All knots are well fitted with a power law. For knot A, when an energy spectra is fitted with a power low, a column density is lower than the 21 cm value. Then, a column density is fixed to the 21 cm value and energy spectra is well fitted with a power law.

4.3 all data set

Fitting results are shown in table 7 and 8. For nucleus of two kinds of pile up contamination, energy spectra is well fitted with power law. For HST-1 of two kinds of pileup contamination, when an energy spectra is fitted with a power law, the column density is lower than the 21 cm value. Therefore, a column density is fixed to the 21 cm value, and energy spectra is fitted with a power law. For HST-1 with a 10% pile up contamination, a chance probability of power law is $4.15 \times 10^{-5}$(3.9 sigma). There is a soft excess. Therefore, an apec model is tried to add to a power law and a reasonable result is obtained. The temperature of 0.2 keV and metal abundance of 0.00 are obtained. In an apec model, a relative abundance is fixed to a value of Grevesse & Anders (1989). A relative abundance is cannot be free because of large uncertainty. A power low model for HST-1 cannot be rejected statistically. For knot A, when an energy spectra is fitted with a power law, the column density is lower than the 21 cm value. Therefore, a column density is fixed to the 21 cm value and energy spectra is fitted with a power law. There is a soft excess as shown in figure 2 (top). The chance probability of a power law is $1.9 \times 10^{-21}$. A power law model is rejected statistically. Therefore, an apec model is added to a power law and a reasonable result is obtained as shown in figure 2 (bottom). The temperature of 0.2 keV and metal abundance of 0.00 are obtained. The flux ratio of an apec model to total is 10%. In an apec model, a relative abundance is fixed to a value of Grevesse & Anders (1989). A relative abundance is cannot be free because of large uncertainty. The metal abundance is low compared with hot gas of Virgo cluster which metal abundance is 1 solar (Belsole et al 2001). When a metal abundance is fixed to 0.1 solar or 1 solar, reasonable results are obtained as shown in table 9.

4.4 comparison between old calibration data and new calibration data

The comparison of fitting parameters with Osone(2017) is shown in table 10. In Osone(2017), data is selected with less than 5% pile up contamination by PIMMS tool. As fitting result with new calibration data for a comparison, selected data with less than 5% pile up contamination by pileup_map command is used.

For former data set, incorrectness of old calibration data is unknown. The 1 keV flux of power law is 40% for HST-1 and 20% for knot A, lower than that with old calibration data, respectively.

For latter data set, for nucleus and nucleus including HST-1, with old calibration data, when an energy spectra is fitted with a power law, there is an excess above 7 keV. A chance probability of a power law is 0.0030 for nucleus and $4.5 \times 10^{-10}$ for nucleus including HST-1, respectively. However, with new calibration model, an energy spectra for nucleus is well fitted with a power law. Therefore, hard excess is due to an incorrectness of old calibration data. For knot A, 1 keV flux of power law is 20% lower than that with old calibration data.

For all data set, for nucleus and nucleus including HST-1, with old calibration data, when energy spectra is fitted with power law, there is an excess above 7 keV. A chance probability of a power law is 0.0014 for nucleus and $2.7 \times 10^{-10}$ for nucleus including HST-1, respectively. However, with new calibration data, an energy spectra for nucleus is well fitted with a power law. Therefore, a hard excess is due to an incorrectness of old calibration data. For HST-1, 1 keV flux of power law is 22% lower than that with old calibration data. For knot A, with old calibration data, when energy spectra is fitted with a combination model of power law and apec, a high temperature of 7.41 keV and a high photon index of 3 are obtained. However, with new calibration data, the temperature of 0.2 keV and a photon index of 2 are obtained. Therefore, abnormal values are due to an incorrectness of old calibration data.

5.discussion

5.1 temperature of gas

Dainotti et al. (2012) pointed out the decreasing soft X-ray emission in the surroundings from knot E to knot F as a cosmic ray cocoon. It is considered that a soft X-ray is absorbed in the compressed interstellar medium of the surroundings. The detection of a thermal emission for the knot A is possible with this argument. The abundance of a thermal emission is low compared with hot gas of Virgo cluster which metal abundance is 1 solar

(Belsole et al 2001). This may suggests that the heating by a shock is occurred inside jet, not outside jet.

The ratio of a temperature is given by $T_2/T_1 = 2\gamma(\gamma-1)M^2/(\gamma+1)^2$ for a strong shock $M \gg 1$ (Longair 1992). Here, $T_1$ is a temperature outside jet and $T_2$ is that inside jet. $M$ is a mach number of a shock wave. $\gamma$ is a ratio of a specific heat. When $\gamma$ is given as 5/3 for monatomic gas, $T_2/T_1 = 0.3\ M^2 \gg 1$. It is possible that the gas in the jet is heated by a shock and it is observed as thermal emission. The ratio of a pressure is given by $p_2/p_1 = 2\gamma M^2/(\gamma+1)$ for a strong shock $M \gg 1$ (Longair 1992). Here, $p_1$ is a pressure outside jet and $p_2$ is a pressure inside jet. When $\gamma$ is given as 5/3 for monatomic gas, $p_2/p_1 = 1.25\ M^2 \gg 1$. There is no pressure valance between inside jet and outside jet.

5.2 plasma density

A normalization of an apec model is given as $10^{-14} \times n_e n_i V / 4\pi D_A^2 (1+z)^2$. Here, $n_e$ is an electron density in units of cm$^{-3}$ and $n_i$ is an ion density in units of cm$^{-3}$. $V$ is a volume of source region in units of cm$^3$. $D_A$ is an angular diameter distance to M87 in units of cm. $z$ is a redshift. An electron density $n_e$ is assumed to be equal to an ion density $n_i$ and an ion density $n_i$ is derived as 11.8 cm$^{-3}$.

The rotation measure for knot A is RM~200 rad m$^{-2}$ (Algaba, Asada, and Nakamura 2016). RM is given as $8.12 \times 10^3\ n_e\ B\ L$ (Longair 1992). Here, $n_e$ is a plasma density in units of m$^{-3}$, $B$ is a magnetic field in units of T and $L$ is a size of a region in units of pc. A magnetic field for HST-1 is estimated to be 0.6 mG(Harris et al. 2009). Hence, the magnetic field for knot A is assumed to be 1 mG. A plasma density is given as $1.6 \times 10^{-3}$ cm$^{-3}$ with a magnetic field of 1 mG and a size of 156 pc for knot A. This value differs from an X-ray energy spectra analysis by an order of 4.

The lower limit of 90% confidence level statistical error of a plasma density from X-ray energy spectra is 10.5 cm$^{-3}$. When a metal abundance is fixed to 0.1 or 1 solar, plasma density is 4.7 cm$^{-3}$ for 0.1 solar and 1.6 cm$^{-3}$ for 1 solar, respectively.

The ratio of a density is given by $\rho_2/\rho_1 = (\gamma+1)/(\gamma-1)$ for a strong shock $M \gg 1$ (Longair 1992). Here, $\rho_1$ is a material density outside jet and $\rho_2$ is that inside jet. When $\gamma$ is given as 5/3 for monatomic gas, $\rho_2/\rho_1 = 4$. It is possible that a plasma density is comparable with a neutral density of a few cm$^{-3}$. For SS433, there is an interaction between a jet and an interstellar medium along a jet and a monocular medium CO is observed with a density of 3 cm$^{-3}$ (Yamamoto et al. 2008). Further study is needed for the difference between a plasma density from Xray energy spectra analysis and that from a rotation measure.

5.3 origin of jet

In an apec model, a relative abundance is fixed to Grevesse & Anders(1989). A relative abundance cannot be set free by a large uncertainty. The metal abundance of 1 solar which is same with hot gas of Virgo cluster cannot be rejected statistically. Best fitting parameter of metal abundance of 0.00 may mean only proton and electron pair as a composition of a jet.

One of the origin of an extreme high energy cosmic ray is considered to be a jet in an active galaxy. As a chemical composition of an extreme high energy cosmic ray, Telescope Array (TA) data suggests a proton at all energy bin and cannot decide elements by an uncertainty at log E>19.0 (Abbasi et al., 2018). Auger data suggests a transition from proton to heavy nucleus as an energy of cosmic ray increase (Neto et al., 2020). An arrival direction of a joint data of TA and Auger show a correlation with starburst galaxies with a 4.7 sigma significance (Matteo et al 2023). With a galactic magnetic field and an extra galactic magnetic field, M87 is a possible source of an extreme high energy cosmic ray accelerated during a flare about 10-12 Myr ago (Olen et al. 2019). However, a heavy nucleus is needed in this model. Therefore, M87 may not be source of an extreme high energy cosmic ray.

Acknowledgement

I thank *Chandra* archive data center for archive data and CIAO software and CXC help for kind support.

Table1. The position of source regions for obsID 18232. 1" is 78 pc.

| Name | RA | DEC | Distance from nucleus | radius of Region |
|---|---|---|---|---|
| nucleus | $12^h30^m49^s.46$ | $12°\ 23'\ 27".9$ |  | 0".5 |
| HST-1 | $12^h30^m49^s.38$ | $12°\ 23'\ 28".4$ | 1".2 (94 pc) | 0".6 |
| D | $12^h30^m49^s.27$ | $12°\ 23'\ 28".9$ | 2".9 (226 pc) | 0".75 |
| A | $12^h30^m48^s.65$ | $12°\ 23'\ 32".3$ | 12".8 (998 pc) | 1".0 |

Table2-1. The used observation list for nucleus which is less than 5% pile up contamination.

| obsID | PI | obs date | number |
|---|---|---|---|
| former data set |  |  |  |
| 16042 |  | 2013.12 | 1 |
| 17056~17057 |  | 2014.12~2015.3 | 2 |
| exposure time | 14ks |  |  |
| latter data set |  |  |  |
| 18809~18813 | Cheng | 2016.3 | 5 |
| 18232~18233 | Russell | 2016.2~2016.4 | 2 |
| 18781~18783 |  | 2016.2~2016.4 | 3 |
| 18836~18838 |  | 2016.4 | 3 |
| 18856 |  | 2016.6 | 1 |
| 20034 | Neilsen | 2017.4 | 1 |
| 19457~19458 | Wong | 2017.2 | 2 |
| exposure time | 345ks |  |  |
| total exposure time | 358ks |  |  |

Table2-2. The used observation list for nucleus which is less than 10% pile up contamination.

| obsID | PI | obs date | number |
|---|---|---|---|
| former data set | | | |
| 3084 | Harris | 2002.2 | 1 |
| 3976 | Harris | 2002.12 | 1 |
| 6303 | Biretta | 2006.5 | 1 |
| 8581 | Biretta | 2008.8 | 1 |
| 10282~10287 | Harris | 2008.11~2009.6 | 6 |
| 11513~11520 | Harris | 2010.4~2010.5 | 8 |
| 13964 | Harris | 2011.12 | 1 |
| 14973~14974 | | 2012.12~2013.3 | 2 |
| 16042 | | 2013.12 | 1 |
| 17056~17057 | | 2014.12~2015.3 | 2 |
| exposure time | 111ks | | |
| latter data set | | | |
| 18809~18813 | Cheng | 2016.3 | 5 |
| 18232~18233 | Russell | 2016.2~2016.4 | 2 |
| 18781~18783 | | 2016.2~2016.4 | 3 |
| 18836~18838 | | 2016.4 | 3 |
| 18856 | | 2016.6 | 1 |
| 20034 | Neilsen | 2017.4 | 1 |
| 19457~19458 | Wong | 2017.2 | 2 |
| exposure time | 345ks | | |
| total exposure time | 456ks | | |

Table2-3. The used observation list for HST-1which is less than 5% pile up contamination.

| obsID | PI | obs date | number |
|---|---|---|---|
| former data set | | | |
| 11518 | Harris | 2010.5 | 1 |
| 13964 | Harris | 2011.12 | 1 |
| 14973~14974 | | 2012.12~2013.3 | 2 |
| 16042 | | 2013.12 | 1 |
| 17056~17057 | | 2014.12~2015.3 | 2 |
| exposure time | 32ks | | |
| latter data set | | | |
| 18809~18813 | Cheng | 2016.3 | 5 |
| 18232~18233 | Russell | 2016.2~2016.4 | 2 |
| 18781~18783 | | 2016.2~2016.4 | 3 |
| 18836~18838 | | 2016.4 | 3 |
| 18856 | | 2016.6 | 1 |
| 20034 | Neilsen | 2017.4 | 1 |
| 19457~19458 | Wong | 2017.2 | 2 |
| exposure time | 345ks | | |
| total exposure time | 377ks | | |

Table2-4. The used observation list for HST-1 which is less than 10% pile up contamination.

| obsID | PI | obs date | number |
|---|---|---|---|
| former data set | | | |
| 3084 3086 | Harris | 2002.2~2002.3 | 2 |
| 10282~10288 | Harris | 2008.11~2009.12 | 7 |
| 11512~11520 | Harris | 2010.4~2010.5 | 9 |
| 13964 | Harris | 2011.12 | 1 |
| 14973~14974 | | 2012.12~2013.3 | 2 |
| 16042 | | 2013.12 | 1 |
| 17056~17057 | | 2014.12~2015.3 | 2 |
| exposure time | 111ks | | |
| latter data set | | | |
| 18809~18813 | Cheng | 2016.3 | 5 |
| 18232~18233 | Russell | 2016.2~2016.4 | 2 |
| 18781~18783 | | 2016.2~2016.4 | 3 |
| 18836~18838 | | 2016.4 | 3 |
| 18856 | | 2016.6 | 1 |
| 20034 | Neilsen | 2017.4 | 1 |
| 19457~19458 | Wong | 2017.2 | 2 |
| exposure time | 345ks | | |
| total exposure time | 456ks | | |

Table2-5. The used observation list for knot D which is less than 5% pile up contamination.

| obsID | PI | obs date | number |
|---|---|---|---|
| former data set | | | |
| 1808 | Wilson | 2000.7 | 1 |
| 3084~3088 | Harris | 2002.2~2002.7 | 5 |
| 3975~3982 | Harris | 2002.11~2003.8 | 8 |
| 4917~4919 | Birreta | 2003.11~2004.8 | 3 |
| 6301~6303,6305 | Birreta | 2006.2~2006.8 | 4 |
| 7352 | Birreta | 2007.5 | 1 |
| 8510~8517 | Harris | 2007.2~2007.3 | 8 |
| 8575~8581 | Birreta | 2007.11~2008.8 | 7 |
| 10282~10288 | Harris | 2008.11~2009.12 | 7 |
| 11512~11520 | Harris | 2010.4~2010.5 | 9 |
| 13964~13965 | Harris | 2011.12~2012.2 | 2 |
| 14973~14974 | | 2012.12~2013.3 | 2 |
| 16042~16043 | | 2013.12~2014.4 | 2 |
| 17056~17057 | | 2014.12~2015.3 | 2 |
| exposure time | 294ks | | |
| latter data set | | | |
| 18809~18813 | Cheng | 2016.3 | 5 |
| 18232~18233 | Russell | 2016.2~2016.4 | 2 |
| 18781~18783 | | 2016.2~2016.4 | 3 |
| 18836~18838 | | 2016.4 | 3 |
| 18856 | | 2016.6 | 1 |
| 20034~20035 | Neilsen | 2017.4 | 2 |
| 19457~19458 | Wong | 2017.2 | 2 |
| 21075~21076 | | 2018.4 | 2 |
| 20488~20489 | Cheng | 2018.1~2018.3 | 2 |
| exposure time | 385ks | | |
| total exposure time | 679ks | | |

Table2-6. The used observation list for knot A which is less than 5% pile up contamination.

| obsID | PI | obs date | number |
|---|---|---|---|
| former data set | | | |
| 1808 | Wilson | 2000.7 | 1 |
| 3084~3088 | Harris | 2002.2~2002.7 | 5 |
| 3975~3982 | Harris | 2002.11~2003.8 | 8 |
| 4917~4919 | Birreta | 2003.11~2004.2 | 3 |
| 5740 | Birreta | 2005.4 | 1 |
| 6301~6303,6305 | Birreta | 2006.2~2006.8 | 4 |
| 7351 7352 | Birreta | 2007.3~2007.5 | 2 |
| 8510~8517 | Harris | 2007.2~2007.3 | 8 |
| 8575~8581 | Birreta | 2007.11~2008.8 | 7 |
| 10282~10288 | Harris | 2008.11~2009.12 | 7 |
| 11512~11520 | Harris | 2010.4~2010.5 | 9 |
| 13964~13965 | Harris | 2011.12~2012.2 | 2 |
| 14973~14974 | | 2012.12~2013.3 | 2 |
| 16042~16043 | | 2013.12~2014.4 | 2 |
| 17056~17057 | | 2014.12~2015.3 | 2 |
| exposure time | 304ks | | |
| latter data set | | | |
| 18809~18813 | Cheng | 2016.3 | 5 |
| 18232~18233 | Russell | 2016.2~2016.4 | 2 |
| 18781~18783 | | 2016.2~2016.4 | 3 |
| 18836~18838 | | 2016.4 | 3 |
| 18856 | | 2016.6 | 1 |
| 20034~20035 | Neilsen | 2017.4 | 2 |
| 19457~19458 | Wong | 2017.2 | 2 |
| 21075~21076 | | 2018.4 | 2 |
| 20488~20489 | Cheng | 2018.1~2018.3 | 2 |
| exposure time | 385ks | | |
| total exposure time | 689ks | | |

Table3. The fitting parameter with a power law model for a former data set. A column density is set free. An error is 90% confidence level statistical error.

|  | nucleus | nucleus(10%) | HST-1 |
|---|---|---|---|
| $N_H$(x$10^{20}$ cm$^{-2}$) | $6.31^{+4.29}_{-4.03}$ | $6.00^{+1.01}_{-0.99}$ | $1.79^{+3.18}_{-1.79}$ |
| photon index | $1.99^{+0.15}_{-0.14}$ | $2.11^{+0.03}_{-0.04}$ | $2.48^{+0.14}_{-0.12}$ |
| 1keV flux(ph/cm$^2$/s/keV) | $4.22^{+0.56}_{-0.49}$ (x$10^{-4}$) | $5.21^{+0.17}_{-0.17}$ (x$10^{-4}$) | $2.27^{+0.25}_{-0.19}$(x$10^{-4}$) |
| $\chi^2$/d.o.f (d.of) | 0.887(101) | 1.034(287) | 1.078(109) |
|  | HST-1(10%) | D | A |
| $N_H$(x$10^{20}$ cm$^{-2}$) | $1.79^{+1.10}_{-1.07}$ | $0.21^{+1.09}_{-0.21}$ | 0.00 |
| photon index | $2.42^{+0.05}_{-0.04}$ | $2.23^{+0.05}_{-0.04}$ | $2.43^{+0.02}_{-0.01}$ |
| 1keV flux(ph/cm$^2$/s/keV) | $4.16^{+0.16}_{-0.15}$(x$10^{-4}$) | $1.11^{+0.05}_{-0.02}$(x$10^{-4}$) | $2.20^{+0.02}_{-0.02}$(x$10^{-4}$) |
| $\chi^2$/d.o.f (d.o.f) | 0.950(223) | 1.004(238) | 1.357(294) |

Table4. The fitting parameter with a column density fixed to the 21 cm value for a former data set. Energy spectra is fitted with a power law or a combination model of a power law and an apec. An error is 90% confidence level statistical error.

| Model |  | D | A |
|---|---|---|---|
| PL | photon index | $2.28^{+0.02}_{-0.02}$ | $2.50^{+0.02}_{-0.02}$ |
|  | 1keV flux(ph/cm$^2$/s/keV) | $1.16^{+0.02}_{-0.01}$ (x$10^{-4}$) | $2.32^{+0.02}_{-0.02}$ (x$10^{-4}$) |
|  | $\chi^2$/d.o.f (d.of) | 1.018(239) | 1.653(295) |
| PL+APEC | photon index |  | $2.29^{+0.04}_{-0.03}$ |
|  | 1 keV flux(ph/cm$^2$/s/keV) |  | $2.01^{+0.06}_{-0.06}$(x$10^{-4}$) |
|  | $kT$(keV) |  | $0.19^{+0.00}_{-0.01}$ |
|  | abundance |  | $0.00^{+0.01}$ |
|  | normalization |  | $3.13^{+0.11}_{-1.01}$(x$10^{-3}$) |
|  | $\chi^2$/d.o.f(d.o.f) |  | 0.955(292) |

Table5. The fitting parameter with a power law for a latter data set. A column density is set free. An error is 90% confidence level statistical error.

|  | nucleus | HST-1 |
|---|---|---|
| $N_H$ (x$10^{20}$ cm$^{-2}$) | $8.66^{+1.20}_{-1.17}$ | $1.17^{+1.69}_{-1.17}$ |
| photon index | $2.19^{+0.03}_{-0.04}$ | $2.30^{+0.06}_{-0.05}$ |
| 1keV flux(ph/cm$^2$/s/keV) | $2.87^{+0.09}_{-0.09}$ (x$10^{-4}$) | $1.19^{+0.06}_{-0.06}$ (x$10^{-4}$) |
| $\chi^2$/d.o.f (d.of) | 1.056(320) | 1.086(228) |
|  | D | A |
| $N_H$ (x$10^{20}$ cm$^{-2}$) | $1.66^{+1.34}_{-1.31}$ | 0.00 |
| photon index | $2.31^{+0.05}_{-0.04}$ | $2.38^{+0.02}_{-0.02}$ |
| 1keV flux(ph/cm$^2$/s/keV) | $1.54^{+0.06}_{-0.06}$(x$10^{-4}$) | $1.95^{+0.03}_{-0.02}$(x$10^{-4}$) |
| $\chi^2$/d.o.f (d.o.f) | 0.984(272) | 1.098(298) |

Table 6. The fitting parameter with a column density fixed to the 21 cm value for a latter data set. An energy spectra is fitted with a power law. An error is 90% confidence level statistical error.

| Model |  | A |
|---|---|---|
| PL | photon index | $2.43^{+0.02}_{-0.02}$ |
|  | 1keV flux(ph/cm$^2$/s/keV) | $2.04^{+0.03}_{-0.03}$ (x$10^{-4}$) |
|  | $\chi^2$/d.o.f (d.of) | 1.209(299) |

Table 7. The fitting parameter with a power law for all data set. A column density is set free. An error is 90% confidence level statistical error.

|  | nucleus | nucleus(10%) | HST-1 |
|---|---|---|---|
| $N_H$(x$10^{20}$ cm$^{-2}$) | $8.26^{+1.14}_{-1.11}$ | $3.95^{+0.74}_{-0.73}$ | $0.00^{+1.03}$ |
| photon index | $2.17^{+0.04}_{-0.03}$ | $2.14^{+0.03}_{-0.02}$ | $2.33^{+0.03}_{-0.03}$ |
| 1keV flux(ph/cm$^2$/s/keV) | $2.93^{+0.09}_{-0.09}$ (x$10^{-4}$) | $3.40^{+0.08}_{-0.08}$ (x$10^{-4}$) | $1.28^{+0.05}_{-0.02}$ (x$10^{-4}$) |
| $\chi^2$/d.o.f (d.of) | 1.061(326) | 1.143(364) | 0.988(241) |
|  | HST-1(10%) | D | A |
| $N_H$(x$10^{20}$ cm$^{-2}$) | $0.00^{+0.10}$ | $2.40^{+0.80}_{-0.80}$ | 0.00 |
| photon index | $2.51^{+0.02}_{-0.02}$ | $2.27^{+0.03}_{-0.03}$ | $2.43^{+0.01}_{-0.02}$ |
| 1keV flux(ph/cm$^2$/s/keV) | $2.18^{+0.02}_{-0.03}$(x$10^{-4}$) | $1.35^{+0.03}_{-0.04}$(x$10^{-4}$) | $2.12^{+0.01}_{-0.02}$(x$10^{-4}$) |
| $\chi^2$/d.o.f (d.o.f) | 1.191(287) | 1.096(315) | 1.485(354) |

Table8 The fitting parameter with a column density fixed to the 21 cm value for all data set. An energy spectra is fitted with a power law or a combination model of a power law and an apec. An error is 90% confidence level statistical error.

| Model |  | HST-1 | HST-1(10%) | A |
|---|---|---|---|---|
| PL | photon index | $2.37^{+0.03}_{-0.03}$ | $2.56^{+0.02}_{-0.02}$ | $2.49^{+0.01}_{-0.02}$ |
|  | 1keV flux(ph/cm$^2$/s/keV) | $1.34^{+0.03}_{-0.02}$ (x$10^{-4}$) | $2.28^{+0.03}_{-0.02}$ (x$10^{-4}$) | $2.22^{+0.02}_{-0.01}$ (x$10^{-4}$) |
|  | $\chi^2$/d.o.f (d.of) | 1.006(242) | 1.362(288) | 1.882(355) |
| PL+APEC | photon index |  | $2.40^{+0.04}_{-0.04}$ | $2.28^{+0.03}_{-0.03}$ |
|  | 1 keV flux(ph/cm$^2$/s/keV) |  | $2.03^{+0.06}_{-0.07}$(x$10^{-4}$) | $1.92^{+0.04}_{-0.03}$(x$10^{-4}$) |
|  | $kT$(keV) |  | $0.22^{+0.02}_{-0.02}$ | $0.20^{+0.01}_{-0.02}$ |
|  | abundance |  | 0.05 | $0.00^{+0.00}$ |
|  | normalization |  | $6.83^{+7.07}_{-6.73}$(x$10^{-4}$) | $2.67^{+0.77}_{-0.63}$(x$10^{-3}$) |
|  | $\chi^2$/d.o.f(d.o.f) |  | 0.942(285) | 0.915(352) |

Table9. The fitting parameter with a combination model of a power law and an apec for knot A of all data set. A column density is fixed to the 21 cm value. A metal abundance is fixed to 0.1 solar or 1 solar. An error is 90% confidence level statistical error.

| abundance 0.1 solar | photon index | $2.33^{+0.02}_{-0.03}$ |
|---|---|---|
| | 1 keV flux(ph/cm$^2$/s/keV) | $2.00^{+0.02}_{-0.03}(\times 10^{-4})$ |
| | $kT$(keV) | $0.20^{+0.01}_{-0.01}$ |
| | normalization | $4.10^{+0.39}_{-0.39}(\times 10^{-4})$ |
| | $\chi^2$/d.o.f(d.o.f) | 1.015(353) |
| abundance 1 solar | photon index | $2.34^{+0.02}_{-0.02}$ |
| | 1 keV flux(ph/cm$^2$/s/keV) | $2.01^{+0.03}_{-0.02}(\times 10^{-4})$ |
| | $kT$(keV) | $0.20^{+0.01}_{-0.01}$ |
| | normalization | $4.71^{+0.44}_{-0.44}(\times 10^{-5})$ |
| | $\chi^2$/d.o.f(d.o.f) | 1.048(353) |

Table10. A comparison of fitting parameters between old calibration data and new calibration data.　0 in table means no change within 90% confidence level statistical error.

|  | nucleus | HST-1 | D | A |
|---|---|---|---|---|
| former data set | | | | |
| column density | 0 | 0 | 0 | |
| photon index | 0 | 0 | 0 | 0 |
| 1 keV flux | 0 | 40%down | 0 | 30% down |
| abundance | | | | 0 |
| kT | | | | 0.04keVdown |
| normalization | | | | 0 |
| latter data set | | | | |
| column density | 0 | 0 | 0 | |
| photon index | 0 | 0 | 0 | 0.11 up |
| 1keV flux | 0 | 0 | 0 | 24%down |
| all data | | | | |
| column density | 0 | 0 | 0 | |
| photon index | 0 | 0 | 0 | 0.69down |
| 1 keV flux | 0 | 25%down | 0 | 17%down |
| abundance | | | | 0 |
| kT | | | | 7.2keVdown |
| normalization | | | | 0 |

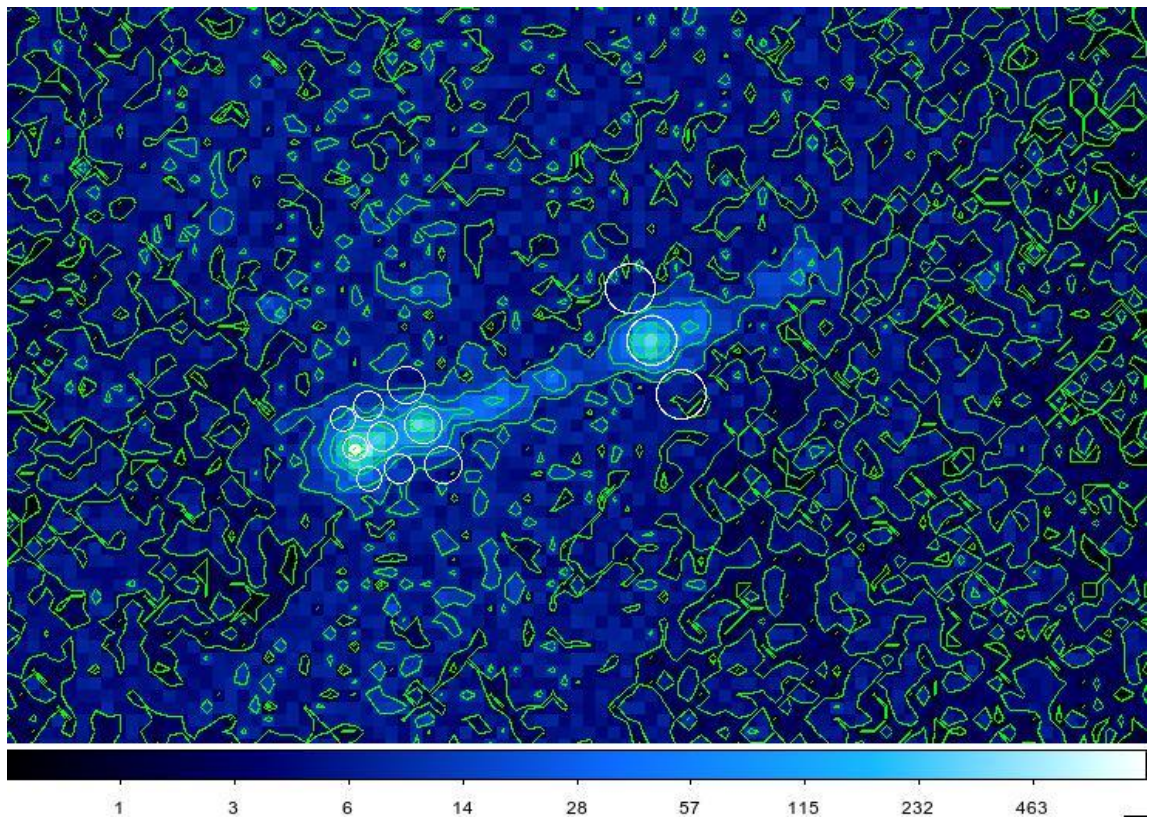

Figure1. ds9 image for obsID 18232. From left, nucleus, HST-1, knot D and knot A are shown in a white circle. Two neighbour regions to each knot as background are shown in a white circle, respectively.

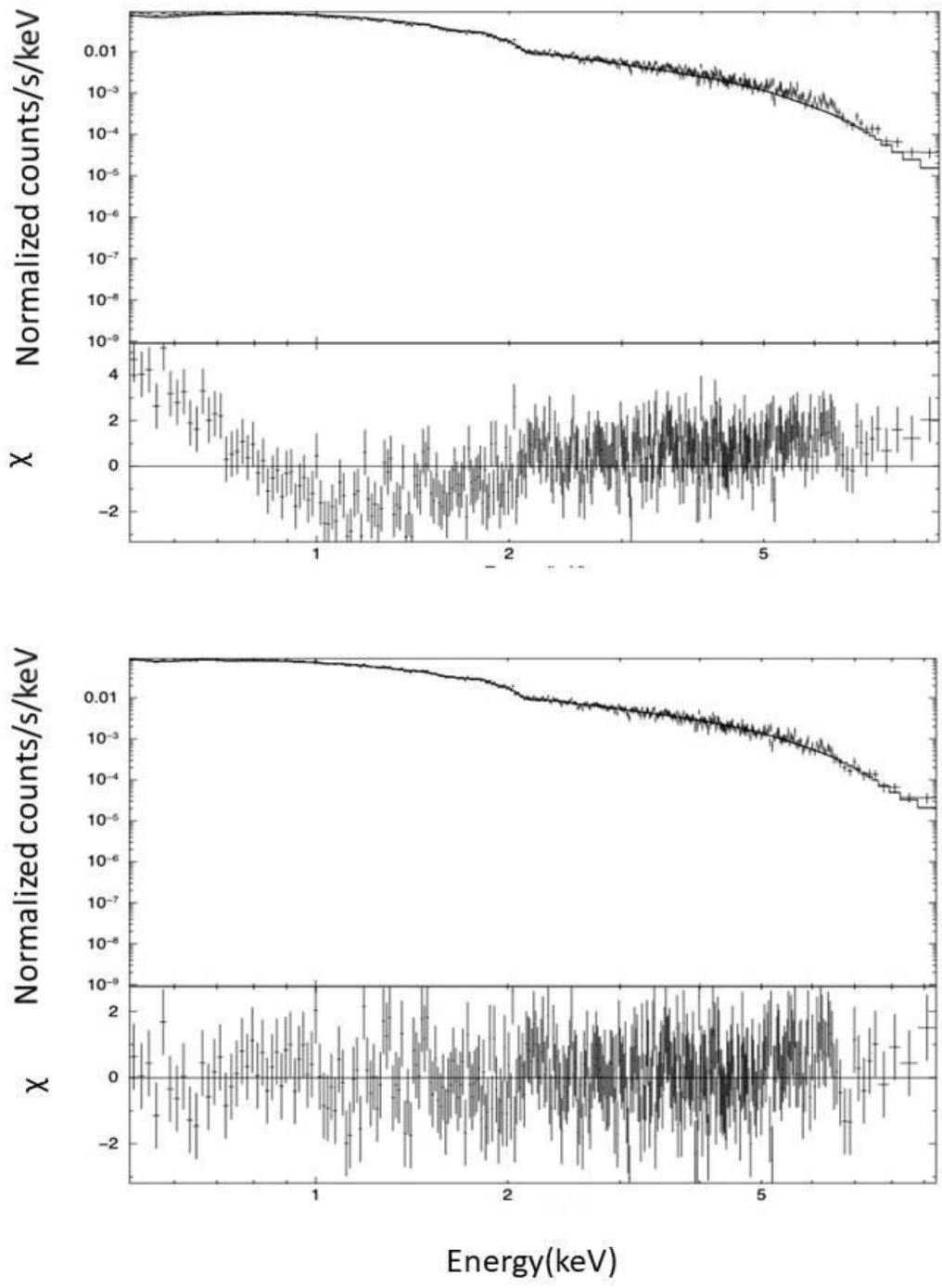

Figure2. The energy spectra fitted with a power law (top) and a combination model of a power law and an apec (bottom) for knot A of all data. Here, a column density is fixed to the 21 cm value.